\documentclass[12pt]{article}
\title{Meromorphic traveling wave solutions of the Kuramoto--Sivashinsky equation}
\author{Alexandre Eremenko\thanks{Supported by NSF grants DMS-0100512 and DMS 0244547.}}
\def\C{{\bf C}}    
\def\bC{{\bf\overline{C}}}
\def\R{{\bf R}}    
\def\Z{{\bf Z}}
\begin{document}
\maketitle
\begin{abstract}
We determine all cases when there exists a meromorphic solution
of the ODE
$$\nu w^{\prime\prime\prime}+bw^{\prime\prime}+\mu w'+w^2/2+A=0.$$
This equation describes traveling waves solutions of
the Kuramoto-Sivashinsky equation.
It turns out that there are no other meromorphic solutions
besides those explicit solutions found by Kuramoto and Kudryashov.
The general method used in this paper, based on Nevanlinna theory,
is applicable to finding all meromorphic solutions of a wide class
of non-linear ODE.

\end{abstract}
The Kuramoto--Sivashinsky equation
$$\phi_t+\nu \phi_{xxxx}+b \phi_{xxx}+\mu \phi_{xx}+\phi\phi_x=0,
\quad \nu,b,\mu\in\R,\quad\nu\neq 0$$
arises in several problems of physics and chemistry
\cite{KuramotoT}, and it was intensively studied
in the recent years \cite{Conte-M,Hone,KuramotoT,Ku1,Ku2,Thual,Conte}.
Solutions of the form of a traveling wave 
$$\phi(x,t)=c+w(z),\quad z=x-ct,$$
satisfy the ordinary differential equation
\begin{equation}\label{1}
\nu w^{\prime\prime\prime}+bw^{\prime\prime}+\mu w'+w^2/2+A=0,
\quad\nu\neq 0,
\end{equation}
which is the object of our study here. We allow complex values for
parameters $\nu,\mu,b$ and $A$ in the equation (\ref{1}). 

It is known \cite{Thual,Conte} that the general
solution of (\ref{1}) has movable
logarithmic branch points, which indicates chaotic behavior.
However, for some values of parameters
$\nu,b,\mu$ and $A$, physically meaningful
one-parametric families of
meromorphic solutions were found in \cite{KuramotoT,Ku1,Ku2}.
Here and in what follows, ``meromorphic function'' means a function
meromorphic in the complex plane $\C$.
In \cite{Conte} the possibility of existence of
other meromorphic solutions, except those
found in \cite{Ku1,Ku2,KuramotoT} is discussed. 
All known meromorphic solutions
of the equation (\ref{1}) are elliptic functions
or their degenerations.
More precisely, let us say that a meromorphic function $f$ belongs to
the class $W$ if $f$ is a rational function of $z$,
or a rational function of $\exp(az),\;a\in \C$,
or an elliptic function.
The letter $W$ is chosen for Weierstrass who proved
that only these functions
can satisfy 
an algebraic addition
theorem.

In this paper we will show that for any choice of parameters,
such that $\nu\neq 0$, all meromorphic solutions
of the equation (\ref{1}) belong to the class $W$. Moreover,
there are no meromorphic solutions except those found in
\cite{KuramotoT,Ku1,Ku2}. 

The crucial fact about (\ref{1}) used here is the
following
\vspace{.1in}

\noindent
{\bf Uniqueness Property}: {\em there is exactly one
formal meromorphic Laurent series with a pole at zero
that satisfies the equation.}
\vspace{.1in}

To check this we substitute the series
\begin{equation}
\label{laurent}
w(z)=\sum_{k=m}^\infty c_kz^k\quad\mbox{with}\quad m<0,\;
c_m\neq 0
\end{equation}
into the equation (\ref{1}),
and obtain $m=-3,\; c_{-3}=120\nu\neq 0$, and
the rest of the coefficients $c_k$ are determined uniquely
(see, for example, \cite{Conte-M,E}). The principal part
of the expansion is
\begin{equation}
\label{2}
w(z)=120\nu z^{-3}-15b z^{-2}+\left(\frac{60\mu}{19}-\frac{15b^2}{76\nu}\right)z^{-1}+
\ldots.
\end{equation}

\noindent
{\bf Theorem 1.} {\em All meromorphic solutions $w$ of the equation $(\ref{1})$
belong to the class $W$. If for some values of parameters
such solution $w$ exists, then all other meromorphic solutions
form a one-parametric family $w(z-z_0),\; z_0\in\C$.
Furthermore,
\newline
(i) Elliptic solutions exist only if $b^2=16\mu\nu$.
They are of order $3$ and have one triple pole per parallelogram
of periods.
\newline
(ii) All exponential solutions have the form $P(\tan kz)$, where $P$ is
a polynomial of degree at most $3$ and $k\in\C$.
\newline
(iii) Non-constant rational solutions occur if and only if $b=\mu=A=0$ and
they have the form $w(z)=120\nu(z-z_0)^{-3},\; z_0\in\C.$}
\vspace{.1in}

Statements (i)-(iii) permit to find all values of parameters when 
meromorphic solutions occur, as well as solutions themselves,
explicitly.
It turns out that there are no other elliptic solutions except those
found
by Kudryashov in \cite{Ku2} (see also \cite{Conte}). This fact was recently
independently established by Hone \cite{Hone}. Similarly,
it follows from (ii) that there are no
other exponential solutions except those found
by Kuramoto--Tsuzuki \cite{KuramotoT}
and Kudryashov \cite{Ku1}.

Our Theorem 1 does not exclude the existence of other ``explicit''
solutions, but it implies that all solutions except those
lited in (i)-(iii) have more complicated singularities,
other than poles, like branching points, or essential
isolated singularities in $\C$, or non-isolated singularities.

We will see that the proof of Theorem 1 is of very general
character, and applies to many other equations which have the
uniqueness property of formal Laurent solutions stated above.
In \cite{E} the author proved
a similar result about the generalized Briot--Bouquet equation
$F(w^{(k)},w)=0$, where $F$ is a polynomial in two variables
and $k$ is odd. If $k$ is even, the equation does not have 
the uniqueness property, as stated above. However, the conjecture
that all meromorphic solutions of all generalized Briot--Bouquet
equations belong to the class $W$ is plausible,
and recently Tuen Wai Ng informed the author that he made a progress
towards this conjecture.

It is desirable to search
other interesting ODE's with this uniqueness property.
The method proposed here will permit to find all
their meromorphic solutions.
We also mention that for any given algebraic ODE,
the uniqueness property can be checked with an efficient
algorithm explained in \cite{Co}. 

The proof of Theorem 1 can be based on any of the two standard tools
of analytic theory of differential equations,
Nevanlinna theory or Wiman--Valiron
theory, see \cite[Chap. V]{Wittich} and \cite[Chap. VI]{Erem}. We choose Nevanlinna theory
here as a more general method. For convenience of a reader unfamiliar
with this theory we include the appendix with definitions and
statements of the results we use.
\vspace{.1in}

{\em Proof of Theorem 1}. We write equation (\ref{1}) as
\begin{equation}
\label{eq}
L(w)=w^2-2A,\quad\mbox{where}\quad L(w)=2(\mu w^{\prime\prime\prime}+
b w^{\prime\prime}+\mu w^{\prime}).
\end{equation}
Let $w$ be a meromorphic solution of (\ref{1}).
The symbols $O$ and $o$ in our formulas refer to asymptotics when
$r\to\infty,\; r\not\in E$, where $E\subset [0,\infty)$ is
a set of finite measure.

We consider two cases.
\vspace{.1in}

{\em Case 1}. $w$ has finitely many poles (possibly none). Then 
the
Nevanlinna characteristic $t(r,L(w))$ can be estimated as follows:
\begin{eqnarray*}
T(r,L(w))=m(r,L(w))+O(\log r)\\
\leq m(r,L(w)/w)+m(r,w)+O(\log r)\\
\leq (1+o(1))T(r,w)+O(\log r),
\end{eqnarray*}
where we used property (\ref{xxx}) and the Lemma on the
Lo\-ga\-rith\-mic
de\-ri\-va\-ti\-ve (see the Appendix) to es\-ti\-ma\-te
$m(r,L(w)/w)$.
On the other hand, $T(r,w^2-2A)=2T(r,w)+O(1)$ (Appendix, (\ref{a}),
(\ref{b})).
So (\ref{eq}) gives
$$T(r,w)=O(\log r),$$
thus $w$ is a rational function.

If $z_0$ and $z_1$ are two poles of $w$ in $\C$ then both
$w(z+z_0)$ and $w(z+z_1)$ are solutions of (\ref{1}) with a pole at zero,
thus $w(z)\equiv w(z-z_1+z_0)$ by the uniqueness property, and we
conclude that $w$ is periodic. This is
a contradiction because the only periodic rational functions are constants,
and they do not have poles.

If $w$ has one pole in $\C$, then $w(z)=c(z-z_0)^{-3}+P(z),$
where $P$ is a polynomial. Substituting this
to our equation, we conclude that $P=0$, 
$b=\mu=A=0$ and $c=120\nu.$
This gives (iii).
\vspace{.1in}

{\em Case 2}. $w$ has infinitely many poles. Arguing as above we
conclude that for every pair of poles $z_0$ and $z_1$,
the difference $z_0-z_1$ is a period of $w$.
So the set of all poles is of the form $z_0+\Gamma$ where $\Gamma$
is a non-trivial discrete subgroup of $(\C,+)$.
Thus $\Gamma$ is isomorphic
to either $\Z$ or $\Z\times\Z$, and we consider each case separately.

If $\Gamma$ is isomorphic to $\Z\times\Z$ then $w$ is elliptic
and there is exactly one pole per period. From (\ref{2}) we conclude
that all poles are of order $3$. The residues at these
poles should be zero, so we obtain from (\ref{2})
$b^2=16\mu\nu$. This proves (i).

Now we consider the remaining case when
$\Gamma$ is isomorphic to $\Z$.
Then $\C/\Gamma=\C^*=\C\backslash{0}$, and
$w$ is a simply periodic meromorphic function,
so it factors as $R(\exp(az))$, where $R$ is a meromorphic function
in $\C^*$, having exactly one pole in $\C^*$.
Our goal is to prove that $R$ is rational,

Making the change of the independent
variable $\zeta=\exp(az)$ in (\ref{1})
we obtain
\begin{equation}
\label{tri}
a^3\nu\zeta^3 R^{\prime\prime\prime}+
(3a^3\nu+a^2b)\zeta^2R^{\prime\prime}+(a^3\nu+a^2b+a\mu)\zeta R'=
R^2/2-A.
\end{equation}
Now we argue exactly as in Case 1, denoting the left hand side of (\ref{tri}) by
$L(R)$.
As $R$ has only one pole, the Lemma on the Logarithmic Derivative implies
$$T(r,L(R))\leq (1+o(1))T(r,R)+O(\log r),$$
but $T(r,R^2/2-A)=2T(r,R)+O(1)$, so, by (\ref{tri}),
$T(r,R)=O(\log r)$, and thus $R$ has no
essential singularity at $\infty$. Applying the same argument to $R(1/\zeta)$,
we conclude that $R$ has no essential singularity at zero. So $R$ is rational.

Now it is easy to see from (\ref{tri}) that $R$ cannot have a pole
at $\infty$ (if $R(\zeta)\sim c\zeta^d), \;d>0$ then
the right hand side has order $\zeta^{2d}$ while the left hand side 
has order at most $\zeta^d$).
Similar argument shows that $R$ cannot have a pole at zero.

Thus $R$ has only one pole in $\bC$, and this pole has to be of order $3$
by (\ref{2}). So we obtain statement (ii).

This completes the proof.
\vspace{.1in}

\noindent
{\bf Conclusions and generalizations}.
\vspace{.1in}

The method of this paper permits the following generalization.
Consider an algebraic autonomous differential equation
\begin{equation}
\label{ge}
\sum a_jw^{j_0}(w')^{j_1}\ldots(w^{(k)})^{j_k}=0,
\end{equation}
where $j=(j_0,\ldots,j_m)$ is a multi-index, and $a_j$ are
constants. The number $j_0+\ldots+j_k$ is called the degree
of a monomial.
Uniqueness Property can be replaced by the following
\vspace{.1in}

\noindent
{\bf Finiteness Property}. {\em There are only
finitely many formal Laurent series of the form $(\ref{laurent})$
that satisfy the equation}.
\vspace{.1in}

For any given equation, Finiteness Property can be verified
either by substituting to the equation
a Laurent series with undetermined coefficients
or by an algorithm in \cite{Conte-M}.

\noindent
{\bf Theorem 2}. {\em Suppose that $(\ref{ge})$ has the finiteness property,
so that the equation is satisfied by finitely many Laurent series
$\phi_n,\; 1\leq n\leq p$ of the form $(\ref{laurent})$.
If in addition $(\ref{ge})$ has only one monomial of top degree,
then all meromorphic solutions belong to the class $W$. Each
solution is either
\newline
a) an elliptic with at most $p$ poles per parallelogram of periods, or
\newline
b) has the form $R(e^{az})$, where $R$ is a rational function with
at most $p$ poles in $\C^*$, or
\newline
c) is a rational function $R$ with at most $p$ poles in $\C$}.
\vspace{.1in}

Nevanlinna and Wiman--Valiron theories usually give only
necessary conditions for existence of meromorphic solutions
of non-linear ODE. However, sometimes these necessary conditions are
so strong that they permit to find or classify all meromorphic
solutions. For example, all meromorphic solutions of the
differential equations $F(w',w,z)=0,$ where $F$ is a polynomial
and $w=w(z)$ were classified in \cite{Er1,Er2} in this way.

In combination with the Finiteness Property, 
Nevanlinna theory permits to make a strong conclusion
that all meromorphic solutions belong to the class $W$, and
moreover, to give a priori bounds for degrees of these meromorphic
solutions, as in statements (i)-(iii) of our Theorem 1. Having established
such bounds one can plug the solution with indetermined coefficients
into the equation, and find all meromorphic solutions explicitly.
Such computation can be hard, but in principle it can be always done
in finitely many steps.

Other instances known to the author when such method was aplied
successfully are the paper on Briot--Bouquet-type equations \cite{E}
mentioned above,
and \cite{CH} were all meromorphic solutions of the equation
\begin{equation}
\label{5}
w^{\prime\prime}w-(w')^2+aw^{\prime\prime}+bw'+cw+d
\end{equation}
were found. The method of \cite{CH} is a combination of
the Finiteness Property and Wiman--Valiron theory.
Solutions of (\ref{5}) do not have poles, but for generic
parameters the following version of the Finiteness Property
holds: there are at most two holomorphic solutions $w$ with
$w(0)=0$ in
a neighborhood of $0$.
\vspace{.1in}

\noindent
{\bf Appendix}. 
\vspace{.1in}

Good general introductions to Nevanlinna theory can be found in
\cite{Wittich}, which contains
a chapter on analytic theory of differential equations, and \cite[Ch.VI]{Lang}
The modern development is described in \cite{GO,Hayman}.
Nevanlinna's own books are \cite{Ne1,Ne2}. 

Let $f\not\equiv 0$
be a meromorphic function in a punctured neighborhood
of infinity $\{ z:r_0\leq |z|<\infty\}$. 
Let $n(r,f)$ be the counting function of poles, that
is $n(r,f)$ is the number of poles in the ring $r_0\leq |z|\leq r$,
counting multiplicity. We set for $r>r_0$
\begin{equation}
\label{N}
N(r,f)=\int_{r_0}^r\frac{n(t,f)}{t}dt,
\end{equation}
and
$$m(r,f)=\frac{1}{2\pi}\int_{-\pi}^\pi\log^+|f(re^{i\theta})|\,d\theta,$$
where $x^+=\max\{ x,0\}$.
The Nevanlinna characteristic is defined by
$$T(r,f)=m(r,f)+N(r,f).$$
Using another number $r_0$ in the definition of $N(r,f)$ adds
to the characteristic $O(\log r)$ as $r\to\infty$ and we will
see that such summands are negligible when $f$ has an essential
singularity at infinity.

The characteristic $T(r,f)$ is a non-negative function, and
\vspace{.1in}

\noindent
{\bf 1.} If the singularity of $f$ at infinity is essential
then $T(r,f)$ is increasing and
$T(r,f)/\log r\to\infty$ as $r\to\infty$. If the infinite point
is a removable or a pole, we have $T(r,f)=O(\log r)$.
\vspace{.1in}

\noindent
{\bf 2.} The algebraic properties of $T(r,f)$ are similar to the
properties
of the degree of a rational function:
\begin{eqnarray}
T(r,fg)\leq T(r,f)+T(r,g),\label{a}\\
T(r,f^n)=nT(r,f),\label{b}\\
T(r,f+g)\leq T(r,f)+T(r,g)+O(1),\label{c}\\
T(r,1/f)=T(r,f)+O(1).\label{d}
\end{eqnarray}
Here we assume that the same $r_0$ was used in the definition
of $T(r,f)$ and $T(r,g)$.
Properties (\ref{a}--\ref{c}) are elementary and follow from the
similar properties of $N(r,f)$ and $m(r,f)$,
for example,
\begin{equation}
\label{xxx}
m(r,fg)\leq m(r,f)+m(r,g).
\end{equation}
Property (\ref{d}) is the restatement of the Jensen formula,
which is fundamental for the whole subject.
These properties show that $T(r,f)$ can be
considered as a generalization
of the degree of a rational function to
functions of ``infinite degree'',
that is to meromorphic functions which have an
essential singularity at
infinity. For such functions, the ``generalized degree''
$T(r,f)$ is an increasing
function rather than a number.
If $f$ is a rational function and $f(0)\neq\infty$ we can take
$r_0=0$ in the definition of $N(r,f)$. Then it is easy to see
that $T(r,f)=\deg f\log r+O(1).$

For applications to differential equations,
the most important property
is 
\vspace{.1in}

\noindent
{\bf The Lemma on the Logarithmic Derivative}
$$m(r,f'/f)=O(\log T(r,f)+\log r),\quad r\to\infty,\; r\not\in E,$$
where $E$ is some exceptional set of finite length. The term $\log r$
can be omitted if $f$ has no essential singularity at infinity. 
The exceptional set $E$ may indeed occur but it does not hurt in
most applications. From now on all our asymptotic relations
have to be understood with $r\to\infty,\; r\not\in E$. 

As the differentiation increases the orders of poles by a factor at
most $2$, we obtain $N(r,f')\leq 2N(r,f).$ Combined with the
Lemma on the Logarithmic Derivative, and property (\ref{xxx}) this
gives 
\begin{eqnarray*}T(r,f')=N(r,f')+m(r,f')\leq 2N(r,f)+m(r,ff'/f)
\\ \leq
2N(r,f)+m(r,f)+m(r,f'/f)\leq (2+o(1))T(r,f).
\end{eqnarray*}
Thus $T(r,f^{(n)})=O(T(r,f))$.
If $f$ has no poles,
we obtain
$$T(r,f')=m(r,f')=m(r,ff'/f)\leq m(r,f)+m(r,f'/f)\leq (1+o(1))T(r,f),$$
and, by induction,
$$T(r,f^{n})\leq (1+o(1))T(r,f).$$
Finitely many poles contribute $O(\log r)$ to $N(r,f)$,
so for functions with finitely many poles we have
$$T(r,f^{n})\leq (1+o(1))T(r,f)+O(\log r).$$
Similarly, if $L(f)$ is a linear differential polynomial
of $f$ with rational coefficients, and $f$ has finitely many
poles, we obtain
$$T(r,L(f))\leq (1+o(1))T(r,f)+O(\log r).$$

The author thanks Tuen Wai Ng and Robert Conte for bringing to
his attention the
connection between
papers \cite{E} and \cite{Conte}, and stimulating discussions.

{\em Purdue University

150 N University street,

West Lafayette IN 47907-2067

eremenko{@}math.purdue.edu}
\end{document}